\documentclass[aps,twocolumn,pra,preprintnumbers,amsmath,amssymb,superscriptaddress]{revtex4}
\usepackage{graphicx}
\usepackage{color}
\usepackage[T1]{fontenc}
\usepackage[latin9]{inputenc}
\usepackage{esint}

\begin{document}

\preprint{TIT/HEP-647}
\preprint{UTHEP-678}

\title{Exact mass-coupling relation for the homogeneous sine-Gordon model}

\date{\today}

\author{Zolt\'an Bajnok}
\email[]{bajnok.zoltan@wigner.mta.hu} 
\affiliation{{\it MTA Lend\"{u}let Holographic QFT Group, Wigner Research Centre,
H-1525 Budapest 114, P.O.B. 49, Hungary}}

\author{J\'anos Balog}
\email[]{balog.janos@wigner.mta.hu}
\affiliation{{\it MTA Lend\"{u}let Holographic QFT Group, Wigner Research Centre,
H-1525 Budapest 114, P.O.B. 49, Hungary}}

\author{Katsushi Ito}
\email[]{ito@th.phys.titech.ac.jp}
\affiliation{{\it Department of Physics, Tokyo Institute of Technology, Tokyo 152-8551, Japan}}

\author{Yuji Satoh}
\email[]{ysatoh@het.ph.tsukuba.ac.jp}
\affiliation{{\it Institute of Physics, University of Tsukuba, Tsukuba, Ibaraki 305-8571, Japan}}

\author{G\'{a}bor Zsolt T\'{o}th}
\email[]{toth.gabor.zsolt@wigner.mta.hu}
\affiliation{{\it MTA Lend\"{u}let Holographic QFT Group, Wigner Research Centre,
H-1525 Budapest 114, P.O.B. 49, Hungary}}

\begin{abstract}
We derive the exact mass-coupling relation of the simplest multi-scale quantum integrable model,
i.e.,  the homogeneous sine-Gordon model with two mass scales.
The relation is obtained by comparing the perturbed conformal field theory description 
of the model valid at short distances to the large distance bootstrap description based 
on the model's integrability. In particular, we find a differential equation for the relation 
by constructing conserved tensor currents which satisfy a generalization of the $\Theta$ 
sum rule Ward identity. The mass-coupling relation is written in terms of hypergeometric functions.
\end{abstract}

\maketitle


\section{Introduction}

One of the most difficult problems in a quantum field theory is to determine the 
mass-coupling relation i.e. the relation between the renormalized couplings related to the 
Lagrangian definition of the theory and the physical masses. 
Such an exact relation would express for example the dynamically generated nucleon mass 
in the chiral limit of quantum chromodynamics
in units of the perturbative Lambda-parameter $\Lambda$, which is defined 
in, say, the  $\overline{\rm MS}$ scheme. 
The difficulty lies in the fact that the Lagrangian is defined at short distances 
(or ultraviolet --UV-- scale), while the masses are the parameters at large distances 
(or infrared --IR-- scale).

There is one family of models where such a relation 
can be found exactly, namely, two dimensional integrable models.
The mass/$\Lambda$ ratio
was indeed exactly determined
\cite{Hasenfratz:1990zz,Hasenfratz:1990ab} in the non-linear sigma (NLS) model.
To this end, one adds an external field coupled to one of the conserved charges,
calculates the free energy perturbatively on the UV side, and
compares it to the large field expansion
from the Bethe Ansatz integral equation/the thermodynamic Bethe Ansatz (TBA)
equation \cite{Zamolodchikov:1989cf} 
on the IR side.
Later this method was applied to many other models
\cite{Forgacs:1991rs,Forgacs:1991nk,Balog:1992cm,Fateev:1992tk,
Hollowood:1994np,Evans:1994sy,Evans:1994sv}.

In contrast to the NLS model with
marginally relevant perturbations, 
there is also a large class of integrable models which 
can be defined as perturbations of their UV-limiting conformal field theories (CFTs) 
by strictly relevant scaling operators. In this case, coupling constants 
are dimensionful, and  one can show 
\cite{Zamolodchikov:1990bk,Constantinescu:1993ny}
that they are not renormalized in the perturbative CFT scheme
and hence are physical themselves. 
When a model in this class has only one 
perturbing operator, the relation
between the coupling constant and the (lowest) physical mass boils
down to a single proportionality constant. 
This non-trivial constant was determined as well by the method described above
for the sine-Gordon and affine-Toda field theories
and their reductions 
\cite{Zamolodchikov:1995xk,Fateev:1993av}.

A common feature of all these models is that they have only one mass scale. 
In some of these models the particles have a 
non-trivial spectrum but all mass ratios are encoded in the S-matrix:
the UV/IR relation is complete once the 
lowest mass is expressed by $\Lambda$, the coupling,
or some other physical dimensionful parameter related to the Lagrangian.
However, when the models have several independent perturbing operators,
the particle spectrum continuously depends on the couplings and not fixed 
by the S-matrix. In this sense, such models can be called multi-scale,
to which the method in the single-scale case is not applicable,
and hence there are no results for multi-scale mass-coupling relations in the literature.

The aim of this letter is therefore to provide a novel method which can fill this gap. 
Though our method is conceptually more general, 
we focus on a class of multi-scale quantum integrable models with strictly relevant 
perturbations, i.e., the homogenous sine-Gordon (HSG) model
\cite{FernandezPousa:1996hi,FernandezPousa:1997zb,FernandezPousa:1997iu,
Miramontes:1999hx,CastroAlvaredo:1999em,Dorey:2004qc}.
We  present our ideas in particular for its simplest case with two scales.
The mass-coupling relation gives the one-point functions of the perturbing operators,
encoding all the non-perturbative information which is not captured by the CFT perturbation.
Via the gauge/string duality, it is applied to the four-dimensional maximally supersymmetric  
gauge theory at strong coupling, which is one of the recent main subjects 
in field and string theories:
it provides the missing link to derive an analytic expansion 
\cite{Hatsuda:2010cc,Hatsuda:2011ke,Hatsuda:2011jn,Hatsuda:2012pb} 
of the strong-coupling amplitudes \cite{Alday:2007hr}.
These are also our main motivations.
Below, we analyze the model both from the UV and IR side, and 
compare the results to obtain the mass-coupling relation.

\section{UV: perturbed CFT}

The simplest multi-scale HSG model  is the perturbation of the $su(3)_{2}/u(1)^{2}$
coset CFT by its weight-0  adjoint primary fields. Fortunately the coset allows
an equivalent representation in terms of the projected product 
\cite{Crnkovic:1989ug} of the Ising
and the tricritical Ising (TCI) minimal models, providing a handy calculational
basis: $su(3)_{2}/u(1)^{2}\sim\mathcal{M}_{3,4}\otimes\mathcal{M}_{4,5}$,
where $\mathcal{M}_{p,q}$ stands for the minimal model with central charge
$c = 1- 6(p-q)^2/pq$.
The coset chiral algebra is larger  than the Virasoro algebra, thus its diagonal modular invariant
partition function representing the spectrum
decomposes into the product of  Virasoro characters
non-diagonally as $Z=2\chi_{\frac{1}{16}\frac{3}{80}}\bar{\chi}_{\frac{1}{16}\frac{3}{80}}
+2\chi_{\frac{1}{16}\frac{7}{16}}\bar{\chi}_{\frac{1}{16}\frac{7}{16}}+(\chi_{00}
+\chi_{\frac{1}{2}\frac{3}{2}})(\bar{\chi}_{00}+\bar{\chi}_{\frac{1}{2}\frac{3}{2}})
+(\chi_{\frac{1}{2}0}+\chi_{0\frac{3}{2}})(\bar{\chi}_{\frac{1}{2}0}+\bar{\chi}_{0\frac{3}{2}})
+(\chi_{0\frac{1}{10}}+\chi_{\frac{1}{2}\frac{3}{5}})(\bar{\chi}_{0\frac{1}{10}}
+\bar{\chi}_{\frac{1}{2}\frac{3}{5}})+(\chi_{0\frac{3}{5}}
+\chi_{\frac{1}{2}\frac{1}{10}})(\bar{\chi}_{0\frac{3}{5}}+\bar{\chi}_{\frac{1}{2}\frac{1}{10}})$,
where 
$\chi_{hh'}=\chi_{h}^{(1)}\chi_{h'}^{(2)}$ 
refers to the characters 
in the tensor product with $h,h'$ being  the dimension of primaries. 
The chiral algebra can be taken to be the product
of the free fermion algebra generated by $\psi(z)$ of  dimension $1/2$
on the Ising side and the superconformal algebra generated by $L^{(2)}(z),G(z)$
on the TCI part. The full Virasoro field is the sum $L(z)=L^{(1)}(z)+L^{(2)}(z)$,
where the Ising contribution is $L^{(1)}(z)=-(1/2)\psi(z)\partial\psi(z)$.
There are 4 fields of dimension $(3/5,3/5)$, which can be obtained
from $\Phi(z,\bar{z})\equiv\Phi_{1/10,1/10}(z,\bar{z})$ 
by acting with the left and right 
chiral generators:
\begin{equation}
\Phi_{ij}(z,\bar{z})=\psi_{-1/2}^{(i)}\bar{\psi}_{-1/2}^{(j)}\Phi(z,\bar{z}) \, ,
\end{equation}
where, to streamline the notations, we introduced $\psi_{-1/2}^{(1)}=\psi_{-1/2}$
and $\psi_{-1/2}^{(2)}=\sqrt{5}G_{-1/2}$. This ensures the proper
normalization of the operators $\langle\Phi_{ij}\vert\Phi_{kl}\rangle=\delta_{ik}\delta_{jl}$.
The Lagrangian of the HSG theory is defined to be 
\begin{equation}
\mathcal{L}=\mathcal{L}_{CFT}-\lambda_{i}\bar{\lambda}_{j}\Phi_{ij}(z,\bar{z}) \, ,
\end{equation}
where summation is understood for $i=1,2$ and $j=1,2$. Since the
transformations $\lambda_{i}\to\beta\lambda_{i}$ and 
$\bar{\lambda}_{i}\to\beta^{-1}\bar{\lambda}_{i}$
with $\beta$ being constant
do not change the perturbation we have effectively 3 parameters. We
also have further discrete symmetries: The remnant of the $S_{3}$
Weyl symmetry in the coset translates into the $\lambda_{i}\to\omega_{ij}\lambda_{j}$
invariance of the perturbation, where $\omega_{ij}$ 
stands for the rotation by $\pm2\pi/3$ or the reflection $\lambda_{1}\to-\lambda_{1}$.
We have similar independent transformations for the right chiral half.

\section{IR: scattering theory}
\label{sec.IRscatt}

The Hilbert space on the IR side contains the scattering states 
$\vert\theta_{1},\dots,\theta_{n}\rangle_{a_{1}\dots a_{n}}$ of two types of 
particles with masses $m_{1}$ and $m_{2}$ which can take arbitrary
values. Here $\theta_{j}$ is the rapidity of the $j^{th}$ particle
of type $a_{j}$ whose energy is $E=m_{a_{j}}\cosh\theta_{j}$. The
theory is integrable and the two particle scattering matrix contains
one resonance parameter $\sigma$ \cite{Miramontes:1999hx}: 
\begin{equation}
S_{12}(\theta-\sigma)=-S_{21}(\theta+\sigma)=\tanh\frac{1}{2}(\theta-i\frac{\pi}{2}) \, .
\end{equation}
These fermionic particles scatter on themselves trivially: $S_{11}(\theta)=S_{22}(\theta)=-1$.
Our aim is to express the three IR parameters, $m_{1},m_{2}$ and
$\sigma$ in terms of the UV parameters $\lambda_{i}$ and $\bar{\lambda}_{j}$.
Since the UV parameters depend on the choice of the basis for $\Phi_{ij}$
we have to map these operators to their IR counterparts. 
On the IR side operators are characterized by their form factors.
For a local operator $X$, they are denoted by
\begin{equation}
\langle0\vert X\vert\theta_{1},\dots,\theta_{n}\rangle_{a_{1}\dots a_{n}}
  =F_{a_{1},\dots,a_{n}}^{X}(\{\theta_{i}\}) \, .
\end{equation}
These form factors have the structure 
\begin{equation}
F_{a_{1}\dots a_{n}}^{X}(\{\theta_{i}\})
=Q_{a_{1}\dots a_{n}}^{X}(\{x_{i}\})\prod_{j<k}F_{a_{j}a_{k}}
(\theta_{j},\theta_{k}) \, ,
\end{equation}
where $x_{i}=e^{\theta_{i}}$ and the two particle form factors are
\begin{equation}
F_{11}(\theta_{1},\theta_{2})=F_{22}(\theta_{1},\theta_{2})
=-\frac{\sinh\frac{\theta_{1}-\theta_{2}}{2}}{2\pi(x_{1}+x_{2})} 
\end{equation}
and $F_{12}(\theta_{1},\theta_{2})  \equiv f(\theta_{1}-\theta_{2})$, which  is the
minimal solution of the equation $f(\theta)=S_{12}(\theta)f(\theta+2i\pi)$;
see \cite{CastroAlvaredo:2000em} for the details. $F_{21}(\theta_{1},\theta_{2})$
is then $F_{21}(\theta_{1},\theta_{2})=f(\theta_{2}-\theta_{1})/S_{12}(\theta_{2}-\theta_{1})$.
The factors $Q^X_{a_{1}\dots a_{n}}(\{x_{i}\})$ are polynomials in $x_i$ and $1/x_i$.
For the trace of the stress tensor, $\Theta$, they were calculated explicitly
in  \cite{CastroAlvaredo:2000em,CastroAlvaredo:2000nk}
and have the structure
\begin{equation}
Q_{a_{1}\dots a_{n}}^{\Theta}(\{x_{i}\})=P(\{x_{i}\})^{2}q_{a_{1}\dots a_{n}}(\{x_{i}\}) \, ,
\end{equation}
where $P^{2}=P^{+}P^{-}$ and $P^{\pm}=P_{(1)}^{\pm}+P_{(2)}^{\pm}$ 
contain the contributions of each particle type to the lightcone  momenta: 
$P_{(a)}^{\pm}=m_{a}\sum_{j\in\mathrm{type}\, a}x_{j}^{\pm 1}$. 
We
can easily define four local operators $X_{ab}$ by their form factors:
\begin{equation}
\label{Qq}
Q_{a_{1}\dots a_{n}}^{X_{ab}}=P_{(a)}^{+}P_{(b)}^{-}q_{a_{1}\dots a_{n}} \, .
\end{equation}
We analyzed numerically the UV expansion of their two point functions by including
six particles in the form factor expansion and confirmed 
that they all have dimensions $(3/5,3/5)$. Note that these operators
depend on the masses only through the prefactors $P_{(a)}^{\pm}$.
As a consequence, their vacuum expectation values and matrix elements
inherit the same mass-dependence. The IR $X_{ab}$ operators are the
linear combinations of the perturbing UV operators $\Phi_{ij}$,
and in the following we relate the two bases to each other.

\section{UV- IR operator relation}

In relating the UV and IR bases, note that $\Theta$ can be written
in both languages, 
\begin{equation}
\Theta= -\frac{4}{5}
\sum_{i,j}\lambda_{i}\bar{\lambda}_{j}\Phi_{ij}=\sum_{a,b}X_{ab} \, ,
\end{equation}
and its vacuum expectation value is related to the free energy density
as $\mathcal{F}=-\lim_{V\to\infty}\frac{1}{V}\ln Z=\frac{1}{2}\langle\Theta\rangle$.
From the definition of the partition function we can write 
\begin{eqnarray}
\label{delFPsi}
\partial_{i}\mathcal{F} & = & -\langle\Psi_{i}\rangle \, , 
 \quad\Psi_{i}=-\bar{\lambda}_{j}\Phi_{ij} \,  , \nonumber \\
\bar{\partial}_{j}\mathcal{F} & = & -\langle\bar{\Psi}_{j}\rangle \, , 
 \quad\bar{\Psi}_{j}=-\lambda_{i}\Phi_{ij} \, ,
\end{eqnarray}
where $\partial_{i}$ is the shorthand for $\partial/ \partial\lambda_{i}$
and similarly $\bar{\partial}_{j}$ for $\partial/ \partial\bar{\lambda}_{j}$.
Form factor perturbation theory expresses the change in the particle
masses in terms of the diagonal  one 
particle form factors, $F_{aa}^{X}\equiv F_{aa}^{X}(i\pi,0)$,
of the perturbing operator as \cite{Delfino:1996xp} 
\begin{equation}
\label{delmFPsi}
\partial_{i}m_{a}^{2}=-4\pi F_{aa}^{\Psi_{i}} \, , 
\quad\bar{\partial}_{j}m_{a}^{2}=-4\pi F_{aa}^{\bar{\Psi}_{j}} \, .
\end{equation}
The change in the scattering matrix is related to the diagonal two
particle form factors $F_{abab}^{\Psi_{i}}(\theta)
\equiv\lim_{\epsilon\to0}F_{abab}^{\Psi_{i}}(\theta+i\pi,i\pi,\theta+\epsilon,\epsilon)$
as \cite{Delfino:1996xp}
\begin{eqnarray}
\label{FPsidelS}
8\pi^{2}iF_{abab}^{\Psi_{i}}(\theta)=2m_{a}m_{b}\sinh\theta\,\partial_{i}S_{ab}(\theta) \nonumber \\
-\left(\partial_{i}m_{a}^{2}+\partial_{i}m_{b}^{2}
+2\cosh\theta\,\partial_{i}(m_{a}m_{b})\right)\partial_{\theta}S_{ab}(\theta) \, .
\end{eqnarray}
TBA analyses relate the bulk energy density to the mass and resonance
parameters as $\mathcal{F}=\frac{1}{2}m_{1}m_{2}\cosh\sigma$
(see \cite{Hatsuda:2011ke}).

On the IR basis, taking into account the mass dependence of the operators $X_{ab}$,
it implies for the vacuum expectation values that $\langle X_{aa}\rangle=0$
and $\langle X_{12}+X_{21}\rangle= 2\mathcal{F}$. 
The diagonal one particle matrix element of $\Theta$ is normalized with respect to 
the masses as $F_{aa}^{\Theta}(i\pi,0)=\frac{m_{a}^{2}}{2\pi}$,
which implies 
\begin{equation}
\label{FXm}
2\pi F_{aa}^{X_{bc}}=\delta_{ab}\delta_{ac}m_{a}^{2} \, .
\end{equation}
From the explicit form of $q_{a_{1}\dots a_{n}}$ in 
\cite{CastroAlvaredo:2000em,CastroAlvaredo:2000nk}, one can
calculate that
\begin{equation}
\label{FXdelS}
 4\pi^{2}iF_{1212}^{X_{ab}}(\theta)=m_{a}m_{b}e^{(b-a)\theta}\partial_{\theta}S_{12}(\theta) \, .
\end{equation}

Expanding $\Psi_i$ by $X_{ab}$, and comparing (\ref{delmFPsi}) with (\ref{FXm})
and (\ref{FPsidelS}) with (\ref{FXdelS}),
we arrive at the relation
\begin{eqnarray}
\label{PsiX}
\Psi_{i} & = 
  & -X_{11}\,\partial_{i}\ln m_{1}-X_{12}\,\partial_{i}\ln(m_{1}m_{2}e^{-\sigma})^{1/2}
 \nonumber \\
 &  & -X_{22}\,\partial_{i}\ln m_{2}-  X_{21}\,\partial_{i}\ln(m_{1}m_{2}e^{\sigma})^{1/2} \, .
\end{eqnarray}
A similar relation for  $\bar{\Psi}_{i}$ is obtained by replacing $\partial_{i}$ with $\bar{\partial}_{i}$. 
The consistency of $\langle \Psi_i \rangle $ from (\ref{delFPsi}) and (\ref{PsiX}) gives
$\langle X_{12}\rangle=\frac{1}{2}m_{1}m_{2}e^{-\sigma}$
and $\langle X_{21}\rangle=\frac{1}{2}m_{1}m_{2}e^{\sigma}.$
Together with these results, 
we restrict the mass-coupling relation from conservation laws in the following.

\section{UV conserved charges }

In the UV CFT any element of the chiral algebra, $\Lambda(z)$, is
a component of a conserved current: $\bar{\partial}\Lambda(z)=0$.
Once we switch on the perturbation this is no longer true, but we
can systematically calculate the corrections. The leading order formula is 
\begin{equation}
\label{delbLambda}
\bar{\partial}\Lambda(z,\bar{z})=-\lambda_{i}\bar{\lambda}_{j}
\oint_{z}\frac{dw}{2 i}
\Lambda(z)\Phi_{ij}(w,\bar{z}) \, .
\end{equation}
Comparing the dimensions  on the two sides one can show that higher
order terms cannot contribute and the first order formula is actually exact. 

Given (\ref{delbLambda}), conserved currents are found by 
the counting argument \cite{Zamolodchikov:1987zf,Zamolodchikov:1989zs}. 
For example, at the second level we have three operators: the Ising stress tensor
$L^{(1)}(z)$, the TCI one $L^{(2)}(z)$ and the product $L^{(3)}(z)=\psi(z)G(z)$.
By analyzing carefully their 
operator product expansion (OPE) with the perturbing fields, $\Phi_{ij}$,
we find two conservation laws. The first combination is the conservation
of the energy $L=L^{(1)}+L^{(2)}$,  
\begin{equation}
\bar{\partial}L 
= \pi (1-h) \lambda_{i}\partial\Psi_{i} .
\end{equation}
where $h=\frac{3}{5}$ is the chiral conformal dimension of 
the perturbing operators. The conservation of the other combination, 
\begin{equation}
J^{-}=L^{(1)}+\alpha L^{(3)} \, , 
\quad\alpha=\frac{\sqrt{5}}{4}\frac{\lambda_{1}}{\lambda_{2}} \, ,
\end{equation}
 follows from the singular part of the OPE 
 $J^{-}(z)\lambda_{i}\Phi_{ij}(w,\bar{w})=\frac{3}{2}\frac{v_{i}\Phi_{ij}(w,\bar{w})}{(z-w)^{2}}
 +\frac{5}{2}\frac{v_{i}\partial\Phi_{ij}(w,\bar{w})}{(z-w)}$
as 
\begin{equation}
\label{Jconsrv}
\bar{\partial}J^{-}=\partial J^{+}\equiv v_i\partial \Psi_{i} \, ,
\end{equation}
where 
$v_{1}=\frac{\pi}{2} \lambda_{1}$ and $v_{2}=\frac{\pi}{6}\frac{\lambda_{1}^{2}}{\lambda_{2}}$.
We denote the corresponding conserved charge by $Q$. Clearly
we have similar equations for the anti-chiral half, $\bar{J}^{-}$
and $\bar{J}^{+}$. We can also calculate how the charge $Q$ acts
on $\bar{J}^{-}:$ $[Q,\bar{J}^{-}(z,\bar{z})]=-\pi\oint\frac{dw}{2\pi i}J^{-}(w)\bar{v}_{j}\bar{\Psi}_{j}(z,\bar{z})$.
Using the short distance OPEs we obtain
\begin{equation}
\label{QJ}
[Q,\bar{J}^{-}] =  -\frac{5}{2}v_{i}\bar{v}_{j}\partial\Phi_{ij} \, .
\end{equation}

\section{IR conserved charges}

From the two conservation laws for $L$ and for $J^{-}$ it is clear
that they have linear combinations $\tau_{i}$ such that $\Psi_{i}$
satisfies $\partial\Psi_{i}=\bar{\partial}\tau_{i}$ for $i=1,2$,
and similarly for $\bar{\Psi}_{i}$. As a consequence
$F^{\Psi_{i}}\propto P^{+}$ and $F^{\bar{\Psi}_{i}}\propto P^{-}$,
which together with (\ref{Qq}) and (\ref{PsiX}) give the relations
\begin{equation}
\label{partialfact}
\partial_{i}\ln\left(\frac{m_{1}}{m_{2}}e^{-\sigma}\right)=0 \, , 
\quad\bar{\partial}_{i}\ln\left(\frac{m_{1}}{m_{2}}e^{\sigma}\right)=0 \, .
\end{equation}
Now it is advantageous to introduce the parameters
\begin{equation}
   \mu_{a}=\frac{m_{a}}{2}e^{\sigma_{a}} \, , 
   \quad\bar{\mu}_{a}=\frac{m_{a}}{2}e^{-\sigma_{a}} \ .
\end{equation}
All physical combinations depend only on 
the difference of $\sigma_a$, namely, $\sigma=\sigma_{1}-\sigma_{2}$.
The equations above imply that $\mu_{1}/\mu_{2}$ depends only on
$\eta=\lambda_{1}/\lambda_{2}$ and $\bar{\mu}_{1}/\bar{\mu}_{2}$
on $\bar{\eta}=\bar{\lambda}_{1}/\bar{\lambda}_{2}.$ In this notation
$\langle X_{12}\rangle=2\mu_{2}\bar{\mu}_{1}$ and $\langle X_{21}\rangle=2\mu_{1}\bar{\mu}_{2}$. 

The action of the conserved currents and charges on multi-particle states
are found using their forms such as (\ref{PsiX}), (\ref{Jconsrv})
with (\ref{partialfact}) and the relevant form factors given in section \ref{sec.IRscatt}.
The commutator $[Q,\bar{J}^{-}]$ is thus expressed in terms of the IR basis $X_{ab}$.
Comparing the resulting expression to the UV result (\ref{QJ}), 
we can derive the relation
\begin{equation}
\label{PhiX}
\Phi_{ij}=-\frac{4}{5}(\partial_{i}\ln m_{a})(\bar{\partial}_{j}\ln m_{b}) X_{ba} \, .
\end{equation}

\section{Master formula}

Our final ingredient for the mass-coupling relation is the master formula,
which is a generalization of the $\Theta$ sum rule of  \cite{Delfino:1996nf} 
for a conserved spin two current. Let us assume that $Y^{\mu\nu}$ satisfies
$\partial_{\mu}Y^{\mu\nu}=0$ and that $\Psi$ is some scalar operator,
such that the leading term of their conformal OPE is 
$\langle Y^{--}(z)\Psi(0)\rangle=\frac{C(0)}{z^{2}}+\dots$.
By following the calculation that  leads to the $\Theta$ sum rule we obtain 
\begin{equation}
\label{MFormula}
\int d^{2}x\,\langle Y^{+-}(x)\Psi(0)\rangle_c=-\pi C(0) \, ,
\end{equation}
where $\langle \cdot \rangle_c $ stands for the connected part. 
For this we used relativistic invariance to parametrize the two point
function as $\langle Y^{\mu\nu}(x)\Psi(0)\rangle_c
=-x^{\mu}x^{\nu}r^{-4}C(r^{2})+\eta^{\mu\nu}A(r^{2})+\epsilon^{\mu\nu}B(r^{2})$.
The conservation law then leads to $\frac{G}{r^{2}}=\frac{d}{dr^{2}}(C+G)$,
where $G=C+2A+2B$. In massive theories $C(\infty)=G(\infty)=0$ and
a relevant conformal dimension, $\Delta  < 1$, for $\Psi$ implies $G(0)=0$.

Applying these formulas to the stress tensor we recover the $\Theta$
sum rule: $\int d^{2}x\langle\Theta(x)\Psi(0)\rangle_c =-2\Delta\langle\Psi\rangle$.
Since the second tensor index of $Y^{\mu\nu}$ can be regarded as a label of the current,
the formula can be applied to  the other conserved current $J^\mu \sim Y^{\mu -}$.
This leads to a differential equation for the mass-coupling relation.

\begin{figure}[t]
\begin{center}
   \begin{minipage}{0.5\hsize}
     \hspace*{-1mm}
     \includegraphics[width=46mm]{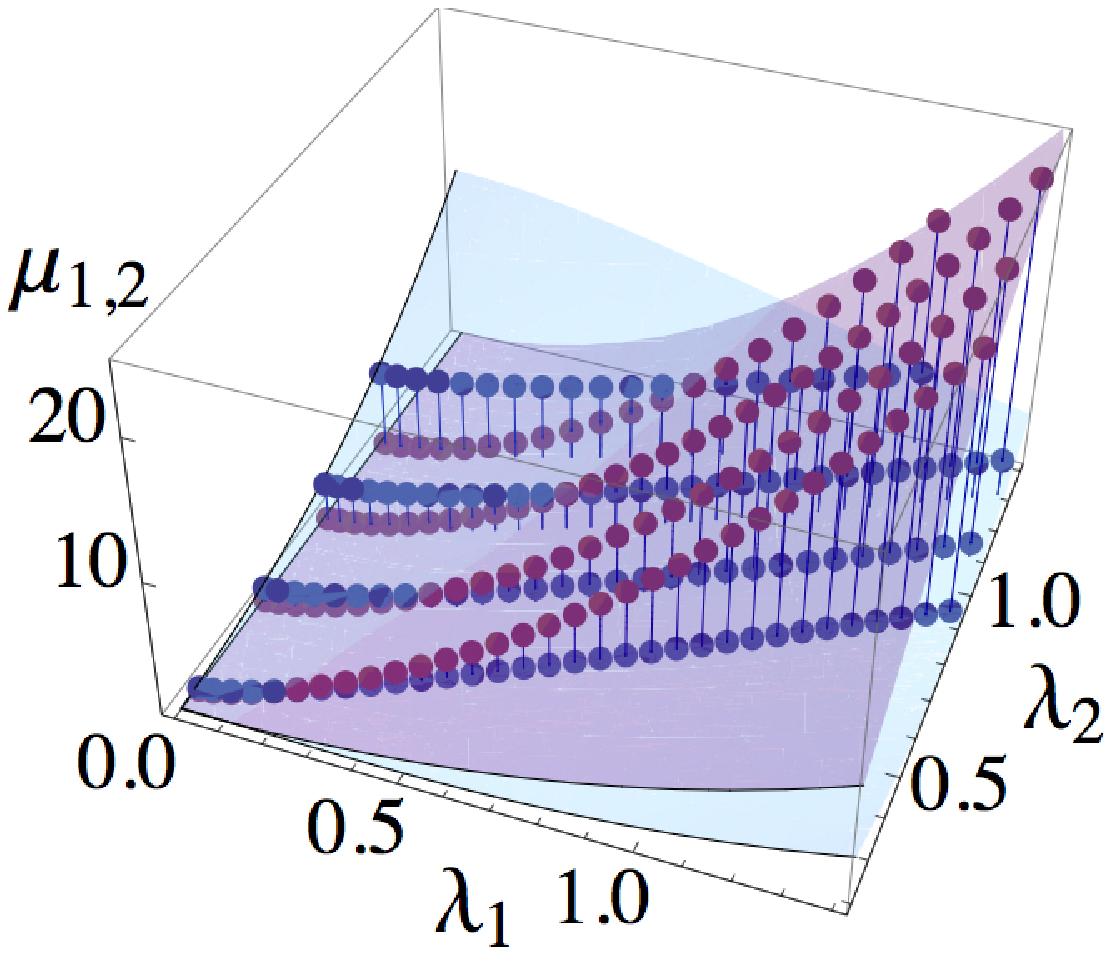}  
   \end{minipage}
   \begin{minipage}{0.48\hsize}
     \hspace*{1mm}
     \vspace*{0mm}
     \includegraphics[width=37mm]{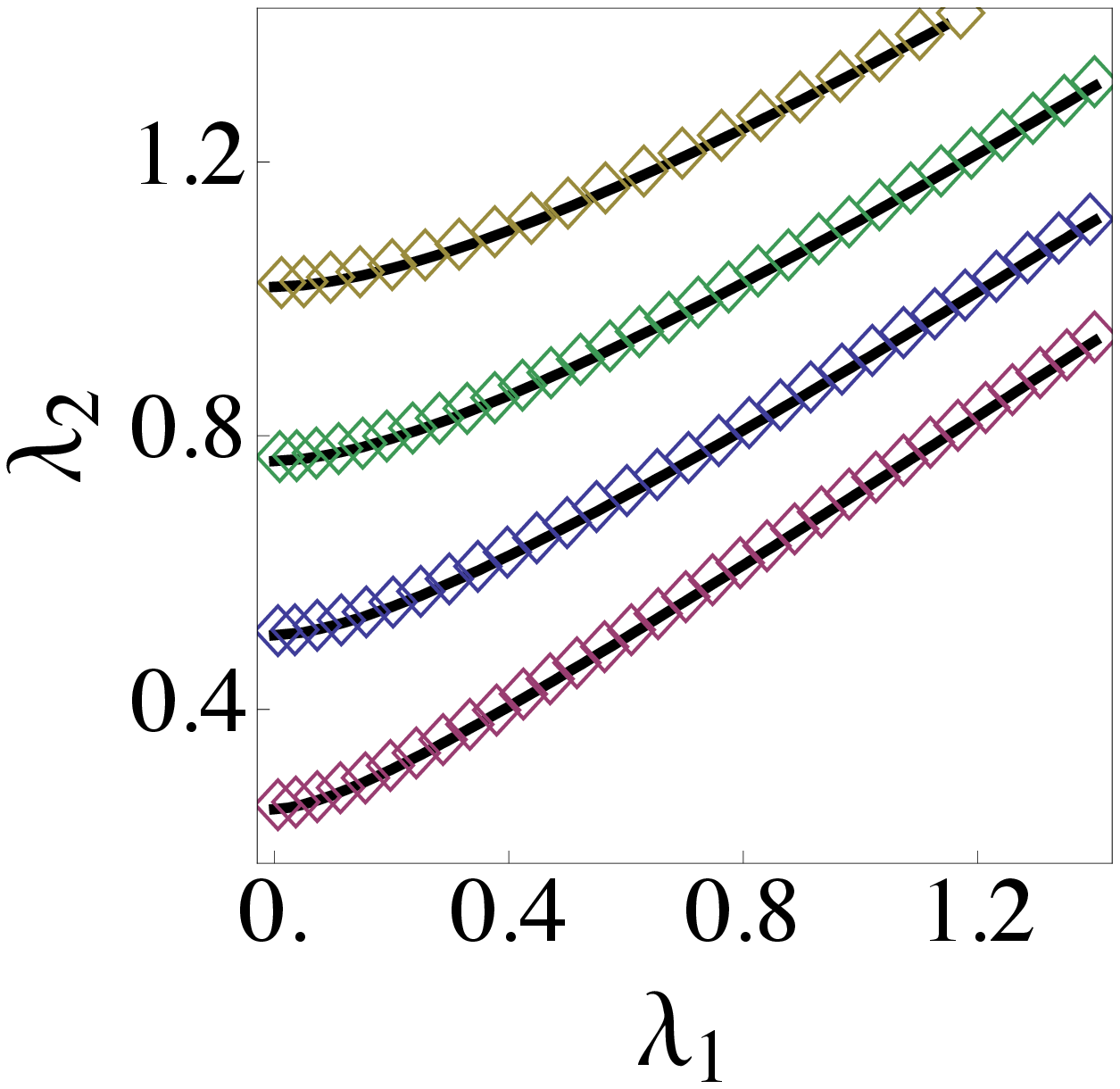} 
   \end{minipage}
 \hfill
\caption{Plots of $(\mu_1, \mu_2)$ versus $(\lambda_1, \lambda_2)$. 
On the left, the red and blue  surfaces represent $\mu_1(\lambda_i)$ 
and $\mu_2(\lambda_i)$ in (\ref{mu1}) and (\ref{mu2}), respectively.  
The red and blue points represent the numerical data 
$( \lambda_1(\mu_a),\lambda_2(\mu_a),\mu_{b})$ $(b=1,2)$ from the TBA equations, 
which are solved  for  given $\mu_a = \bar{\mu}_a$. Each sequence from the bottom 
to the top corresponds to $(\mu_2)^{2/5} = 1/2, 1, 3/2, 2$, with $\mu_1 $ varied. 
$\lambda_i $ are determined by comparing the TBA free energy 
with the CFT perturbation. On the right, the diamonds ($\diamond$) represent 
the projections of the left points to the $(\lambda_1, \lambda_2)$-plane. 
The solid lines are the contours  in the fundamental domain for 
$ \bigl( \mu_2(\lambda_i) \bigr)^{2/5} = 1/2, 1, 3/2, 2$ from (\ref{mu2}).}
\label{fig:fig1}
\end{center}
\end{figure}

\section{Mass-coupling relation}

To see this,  first note that the master formula  (\ref{MFormula}) enables us 
to calculate the free energy Ward identity,
\begin{equation}
\label{delFPhi}
\partial_{i}\bar{\partial}_{j}\mathcal{F}
=-\langle\Phi_{ij}\rangle-\int d^{2}x
      \langle\Psi_{i}(x)\bar{\Psi}_{j}(0)\rangle_c
=-\frac{5}{2}\langle\Phi_{ij}\rangle \, . 
\end{equation}
Together with 
$\mathcal{F}=\mu_{1}\bar{\mu}_{2}+\mu_{2}\bar{\mu}_{1}$, this 
implies complete factorization, i.e.,  
$\mu_{a}$ depends on $\lambda_{i}$ as $\mu_{a}(\lambda_{1},\lambda_{2})$, 
and similarly  
 $\bar{\mu}_a$ as $\bar{\mu}_{a}(\bar{\lambda}_{1},\bar{\lambda}_{2})$.
 This means that the original three-variable mass-coupling relation
is reduced to two identical copies of the chiral two-variable mass-coupling relation.
On dimensional grounds we can thus write 
\begin{equation}
 \mu_{a}=\frac{\lambda_{1}^{5/2}}{2}q_{a}(\eta) \, , 
\quad\bar{\mu}_{a}=\frac{\bar{\lambda}_{1}^{5/2}}{2}q_{a}(\bar{\eta}) \, ,
\end{equation}
 so as to maintain the left-right symmetry of the problem, 
where as before $\eta=\lambda_{1}/\lambda_{2}$.

The master formula implies also that 
\begin{equation}
v_{i}\partial_{i}\langle\Phi_{kj}\rangle
    =\int d^{2}x\langle J^{+}(x)\Phi_{kj}\rangle_c= \frac{\pi}{2} M_{ki}\langle\Phi_{ij}\rangle \, ,
\end{equation}
where from the OPEs we obtain $M_{11}=1$, $M_{12}=M_{21}=\frac{1}{2}\eta$
and $M_{22}=0$. Through (\ref{PhiX}),
this actually translates into the following differential equation for $q_{a}$: 
\begin{equation}
\label{HGeq}
 \eta^{2}\left(1-\frac{\eta^{2}}{3}\right)q_{a}''+\eta\left(4-\frac{2\eta^{2}}{3}\right)q_{a}'
+\frac{5}{4}q_{a}=0
\, ,
\end{equation}
which is a hypergeometric differential equation whose solutions
need to be fixed from the boundary conditions. One special case can
be obtained by sending $\lambda_{1}=\bar{\lambda}_{1}$ to $0$. In
this case only the TCI model is perturbed with $\lambda_{2}\bar{\lambda}_{2}\Phi_{22}$
and the masses are explicitly  known as
$m_{1}=0$ and $m_{2}=\kappa(\lambda_{2}\bar{\lambda}_{2})^{5/4}$ 
with $\kappa = {56(21\pi)^{1/4} \over 5^{5/2} } 
 \bigl({ \Gamma(-\frac{7}{5}) \Gamma(\frac{1}{5}) 
\over \Gamma(\frac{12}{5}) \Gamma(\frac{4}{5})  }  \bigr)^{5/8}$
\cite{Zamolodchikov:1995xk,Hatsuda:2011ke}.
The solution of (\ref{HGeq}) for such vanishing $\mu_1$ is unique up to normalization, giving 
\begin{equation}
\label{mu1}
\mu_{1}(\lambda_{1},\lambda_{2})=B\lambda_{1}^{2}(\lambda_{1}
+\sqrt{3}\lambda_{2})^{1/2}F\biggl(\frac{2\lambda_{1}}{\lambda_{1}+\sqrt{3}\lambda_{2}}\biggr) \, ,
\end{equation}
where $F(z)=\,_{2}F_{1}\left(-\frac{1}{2},\frac{3}{2};3\vert z\right)$.
The $S_3$ symmetry  then yields 
\begin{equation}
\label{mu2}
\mu_{2}(\lambda_{1},\lambda_{2})
=\frac{B}{4}\frac{\left(\sqrt{3}\lambda_{2}-\lambda_{1}\right)^{2}}{(\lambda_{1}+\sqrt{3}\lambda_{2})^{-1/2}}
F\biggl(\frac{\sqrt{3}\lambda_{2}-\lambda_{1}}{\lambda_{1}+\sqrt{3}\lambda_{2}}\biggr) \,.
\end{equation}
(\ref{mu1}) and (\ref{mu2}) hold in the fundamental domain
$0 \leq \lambda_1 \leq \sqrt{3}\lambda_2 $, 
which are  continued outside by the $S_3$ symmetry.
The normalization is fixed by the above single-mass result: $B=\kappa\frac{5\pi}{16\sqrt[4]{3}}$.
This is our main result, which we have checked numerically from the TBA equations 
\cite{CastroAlvaredo:1999em}. FIG. 1 shows the agreement of 
 (\ref{mu1}), (\ref{mu2}), and samples  of numerical data. 
 Furthermore, at $(\lambda_1, \lambda_2) =(\lambda/2, \sqrt{3}\lambda/2)$,
 we confirm that  $\mu_1 = \mu_2 = \frac{B}{2\sqrt{2}} F(1/2) \lambda^{5/2}$,
 which exactly reproduces the mass-coupling relation in the equal-mass case
 \cite{Fateev:1993av,Hatsuda:2011ke}. 
 The mass-coupling relation enables us to express 
 the free energy density $\mathcal{F}$  in terms of $ (\lambda_i,\bar\lambda_i)$, which
 then can be used via (\ref{delFPhi}) to obtain 
 the one-point functions of $\Phi_{ij}$.

\section{Conclusions} 

In this letter we developed a new method to calculate the exact mass-coupling
relation for multi-scale quantum integrable models. We combined form factor
perturbation theory with the construction of conserved tensor currents. The
generalization of the $\Theta$ sum rule Ward identity of these currents
provided a differential equation for the mass coupling relations, leading to solutions 
in terms of hypergeometric functions.
This is the first result for multi-scale mass-coupling relations.
Our work provides the missing link to develop an analytic expansion of 
ten-particle scattering amplitudes of the four-dimensional maximally supersymmetric 
gauge theory at strong coupling around a ${\mathbb Z}_{10}$-symmetric kinematic point
 \footnote{
In \cite{Hatsuda:2011ke}, the mass-coupling relation was studied assuming
that $\lambda_i$ are  polynomials of 
 $\mu_a^{2/5}$.
In  the second order CFT perturbation of ${\cal F}$, that leads to
a  deviation of less than 1 $\%$ from the exact values
 for each contribution from the chiral and anti-chiral sectors.
It is still unclear why such a simple assumption works well effectively.
}.
Although we analyzed here the simplest multi-scale HSG model, 
the methods can 
be extended for other multi-scale perturbed CFTs.
More details and related results will be reported elsewhere \cite{BBIST}.

\begin{acknowledgments}
We would like to thank J. Luis Miramontes for useful conversations.
This work was supported by Japan-Hungary Research Cooperative Program.
Z.~B., J.~B. and  G.~Zs.~T. were supported by a Lend\"ulet Grant and by
OTKA K116505, 
whereas K.~I. and Y.~S. were supported by JSPS Grant-in-Aid for Scientific Research
15K05043 and 24540248.
\end{acknowledgments}



\end{document}